\documentclass[12pt,english]{paper}
\usepackage{amssymb}
\usepackage[T1]{fontenc}
\usepackage[latin1]{inputenc}
\usepackage{graphicx}

\makeatletter

\makeatletter

\makeatletter

\usepackage{geometry}

\makeatletter

\usepackage{epsfig}

\makeatother

\makeatother

\makeatother

\usepackage{babel}
\makeatother
\begin{document}

\title{Evaluation of the neutron background in CsI target for WIMP direct detection when using a reactor neutrino detector as a neutron veto system}

\author{Ye Xu $^a$%
\thanks{Corresponding author, e-mail address: xuy@fjut.edu.cn%
}, Xiangpan Ji $^b$, Haolin Li $^b$, Yulong Feng $^b$}

\maketitle

\begin{flushleft}
$^a$Department of Mathematics and Physics, Fujian University of Technology, Fuzhou 350118, China
\par
$^b$School of Physics, Nankai University, Tianjin 300071, China
\end{flushleft}

\begin{abstract}
A direct WIMP (Weakly Interacting Massive Particle) detector with a
neutron veto system is designed to better reject neutrons. An
experimental configuration is studied in the present paper: a WIMP detectors with CsI(Na) target
 is placed inside a reactor neutrino detector. The neutrino
detector is used as a neutron veto device. The neutron background
for the experimental design has been estimated using the Geant4
simulation. The results show that the neutron background can
decrease to O(0.01) events per year per tonne of CsI(Na). We
calculate the sensitivity to spin-independent WIMP-nucleon elastic
scattering. An exposure of one tonne $\times$ year could reach a
cross-section of about 3$\times$$10^{-11}$ pb.
\end{abstract}


\begin{flushleft}
\end{flushleft}

\section{Introduction}
It is indicated by seven year Wilkinson Microwave Anisotropy Probe
data combined with measurements of baryon acoustic oscillations and
Hubble constant that $\sim$$83\%$ of the matter content in the
Universe is non-baryonic dark matter \cite{WMAP2011, BAQ2010,
H2009}. Weakly Interacting Massive Particles (WIMPs)
\cite{Steigman1985}, predicted by extensions of the Standard Model
of particle physics, are a well-motivated class of candidates for
dark matter. They are distributed in the halo surrounding the Milky
Way. WIMPs may be directly detected through measuring nuclear
recoils in terrestrial detectors produced by their scattering off
target nuclei \cite{Goodman1985, Jungman1996, Gaitskell2004}. The
nuclear reoils is expected to have a roughly exponential energy
distribution with a mean energy in a few tens of keV
\cite{Jungman1996, Bertone2005, Lewin1996}.
\par
In direct searches for WIMPs, there are three different methods used
to detect the nuclear recoils, including collecting ionization,
scintillation and heat signatures induced by them. The background of
this detection is made up of electron recoils produced by $\gamma$
and $\beta$ scattering off electrons, and nuclear recoils produced
by neutrons scattering elastically off target nucleus. It is very
efficient discriminating nuclear recoils from electron recoils with
pulse shape discrimination, hybrid measurements and so on. The
rejection powers of these techniques can even reach  $\>$$10^6$
\cite{CDMSII2010, LAr2008}. For example, the CDMS-II
\cite{CDMSII2010} and EDELWEISS-II \cite{EDELWEISS-II2011}
experiments measure both ionization and heat signatures using
cryogenic germanium detectors in order to discriminate between
nuclear and electron recoils, and the XENON100 \cite{XENON2012} and
ZEPLIN-III \cite{ZEPLIN-III2011} experiments measure both ionization
and scintillation signatures using two-phase xenon detectors.
However, it is very difficult to discriminate between nuclear
recoils induced by WIMPs and by neutrons. In some dark matter experiments and researches,
tagging neutron is applied to reject neutron background. In the ZEPLIN-III experiment, the
$0.5\%$ Gadolinium (Gd) doped polypropylene is used as the neutron
veto device, and its maximum tagging efficiency for neutrons reaches
about $80\%$ \cite{ZEPLIN-III2010}. In Ref. \cite{GdWater2008}, the
$2\%$ Gd-doped water is used as the neutron veto, and its neutron
background can be reduced to 2.2 (1) events per year per tonne of
liquid xenon (liquid argon). In our past work \cite{GdLS2010}, the
reactor neutrino detector with $1\%$ Gd-doped liquid scintillator
(Gd-LS) is used as the neutron veto system, and its neutron
background can be reduced to about 0.3 per year per tonne of liquid
xenon.
\par
Cesium iodide(CsI) crystals as a kind of dark matter target have been applied to
dark matter experiments, such as the KIMS experiment\cite{kims}.
In a dark matter experiment with CsI target, CsI crystals needn't to be cooled with the
liquid nitrogen or xenon, so this detector can be of the simpler structure.
In the Sun, Lu et al.'s work{\cite{slz}}, the rejection power against electron recoil can
reach O($10^7$) with Na-doped CsI crystals. The feasibility of direct WIMPs detection
with a neutron veto based on a neutrino detector had been
validated in our past work\cite{GdLS2010}. So, in the present paper,
a neutrino detector with Gd-LS ($1\%$ Gd-doped) is still used as a
neutron-tagged device and WIMP detectors with CsI target are placed inside the Gd-LS. Here we designed an
experimental configuration: four WIMP detectors with CsI(Na) target are individually placed
inside four reactor neutrino detector modules which are used as a
neutron veto system. The experimental hall of the configuration is
assumed to be located in an underground laboratory with a depth of
910 meter water equivalent (m.w.e.), which is similar to the far
hall in the Daya Bay reactor neutrino
experiment\cite{DayabayProp2007}. Collecting scintillation signals is
considered as the only method of WIMPs detection in our work.
The neutron background for this design are estimated using the
Geant4 \cite{Geant4} simulation.
\par
\section{Detector description}
\begin{figure}
 \centering
 \includegraphics[width=0.7\textwidth]{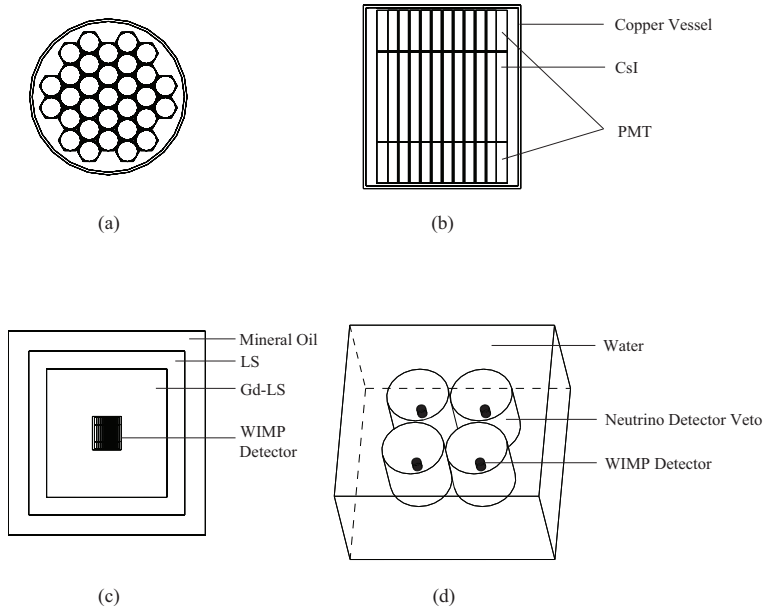}
 \caption{(a) Cross-section for WIMP detector with CsI targets\@. (b) Longitudinal section for WIMP
detector with CsI targets\@. (c) Longitudinal section for a neutrino detector where a WIMP detector is placed inside\@. (d) Four WIMP detectors are
individually placed inside four neutrino detectors in a water
shield\@.}
 \label{fig:detector}
\end{figure}

Four identical WIMP detectors with CsI(Na) target are individually
placed inside four identical neutrino detector modules. The concentration
of Na is 0.02\% in CsI crystals. The
experimental hall of this experimental configuration is assumed to
be located in an underground laboratory with a depth of 910 m.w.e.,
which is similar to the far hall in the Daya Bay reactor neutrino
experiment. The detector is located in a cavern of
20$\times$20$\times$20 $m^3$. The four identical cylindrical
neutrino modules (each 413.6 cm high and 393.6 cm in diameter) are
immersed into a 13$\times$13$\times$8 $m^3$ water pool at a depth of
2.5 meters from the top of the pool and at a distance of 2.5 meters
from each vertical surface of the pool. The detector configuration
is shown in Fig.\ref{fig:detector}.
\par
Each neutrino detector module is partitioned into three enclosed zones. The
innermost zone is filled with Gd-LS, which is surrounded by a zone
filled with unload liquid scintillator (LS). The outermost zone is
filled with transparent mineral oil. 366 8-inch PMTs are mounted in
the mineral oil.
\par
Each WIMP detector (69.6 cm height, 59.9 cm in diameter) consists
of three components: a CsI(Na) crystal array, PMTs and Copper
vessels. The CsI(Na) array is made up of 31 CsI(Na) crystals whose
sections are regular hexagons (34 cm height, 4.5 cm side length) and is
placed inside a 1 cm thick copper vessel (69.2 cm height, 59.5 cm in
diameter) filled with dry nitrogen gas. Fig.\ref{fig:detector} shows
the arrangement of these crystals. 62 3-inch PMTs are individually
mounted on two ends of these crystals. The copper vessel of each WIMP detector is surrounded by an 0.2 cm thick
Aluminum reflector for photons produced in the Gd-LS.
\par
\section{Some features of simulation}
\par
The Geant4 (version 8.2) package\cite{Geant4} has been used in our
simulations. The physics list in the simulations includes
transportation processes, decay processes, low energy processes,
electromagnetic interactions (multiple scattering processes,
ionization processes, scintillation processes, optical processes,
cherenkov processes, Bremsstrahlung processes, etc.) and hadronic
interactions (lepton nuclear processes, fission processes, elastic
scattering processes, inelastic scattering processes, capture
processes, etc.). The hadronic processes include the low energy(<20 GeV), 
high energy(>20GeV) and neutron high-precision(<20 MeV) models.
The cuts for the productions of gammas, electrons
and positrons are 1 mm, 100 $\mu$m and 100$\mu$m, respectively. The
quenching factor is defined as the ratio of the detector response to
nuclear and electron recoils. The Birks factor for protons in the
Gd-LS is set to 0.01 g/cm$^{2}$/MeV, corresponding to the quenching
factor 0.17 at 1 MeV, in our simulation. Besides, We utilize 10 Intel Core i5
CPUs (four cores, each core offer a base of speed of 2.8GHz) in the neutron background evaluation.
\par
\section{Neutron background estimation}
The recoil energies for WIMP interactions with CsI(Na) nuclei were set to
a range from 20 keV to 100 keV in this work. Proton
recoils induced by neutrons and neutron-captured signals are used to
tag neutrons which reach the Gd-LS. The energy deposition produced
by proton recoils is close to a uniform distribution. Neutrons
captured on Gd and H lead to a release of about 8 MeV and 2.2 MeV of
$\gamma$ particles, respectively. Due to the instrumental
limitations of the Gd-LS, we assume neutrons will be tagged if their
energy deposition in the Gd-LS is more than 1 MeV, corresponding to
0.17 MeVee (electron equivalent energy). In the Gd-LS, it is
difficult to distinguish signals induced by neutrons from electron
recoils, which are caused by the radioactivities in the detector
components and the surrounding rocks. But these radioactivities can
be controlled to less than $\sim$ 50 Hz according to the Daya Bay
experiment\cite{DayabayProp2007}. If we assume a 100 $\mu$s for
neutron tagging time window, the indistinguishable signals due to
the radioactivities will result in a total dead time of less than 44
hours per year.
\par
Neutrons are produced from the detector components and their
surrounding rock. For the neutrons from the surrounding rock there
are two origins: first by spontaneous fission and ($\alpha$, n)
reactions due to $\rm{^{238}U}$ and $\rm{^{232}Th}$ in the rock (these neutrons can be omitted
because they are efficiently shielded, see Sec.4.2), and secondly by
cosmic muon interactions with the surrounding rock.
\par
We estimated the number of neutron background in the CsI(Na) target of
one tonne. This number has been normalized to one year of data
taking and are summarized in Tab.\ref{tab:bg}.
\subsection{Neutron background from detector components}
Neutrons from the detector components are induced by ($\alpha$, n)
 reactions due to U and Th. According to Mei et al.\cite{Mei2009}, the
 differential spectra of neutron yield can be expressed as
\begin{center}
 $\displaystyle Y_{i}(E_{n})=N_i{
\sum_{j}\frac{R_{\alpha}(E_{j})}{S_{i}^{m}(E_{j})}}$$\displaystyle
\intop_{0}^{E_{j}}\frac{d\sigma(E_{\alpha},E_{n})}{dE_{\alpha}}dE_{\alpha}$
\end{center}
where $N_i$ is the total number of atoms for the $i^{th}$ element in
the host material, $R_\alpha$$(E_j)$ refers to the $\alpha$-particle
production rate for the decay with the energy $E_j$ from $^{232}Th$
or $^{238}U$ decay chain, $E_{\alpha}$ refers to the $\alpha$
energy, $E_n$ refers to the neutron energy, and $S_{i}^{m}$ is the
mass stopping power of the $i^{th}$ element.
\par
\subsubsection{Neutrons from PMTs in copper vessels}
The U and Th contaminations in the $\rm{SiO_2}$ material are considered
as the only neutron source in the PMTs in our work. Neutrons from
$\rm{SiO_2}$ are emitted with their average energy of 2.68 MeV\cite{Mei2009}.
The PMTs in the copper vessels of the four WIMP
detectors amounts to 248. The U and Th concentrations in the PMT components
can reach ten or even less ppb\cite{MJC}, so a rate of one neutron
emitted per PMT per year is conservatively estimated\cite{AMA}.
Consequently, there are 248 neutrons produced by all the PMTs in the
copper vessels per year. A simulated sample of 2.48$\times$$10^6$ events is used to study this neutron background. 
These events generated isotropically are uniformly distributed in the $\rm{SiO_2}$ material of the PMTs.
\par
The simulation result is summarized in Tab. 1. 2.9 neutron
events/(ton$\cdot$yr) reach the CsI(Na) target and their energy
deposition falls in the same range as that of the WIMP interactions, as
seen in Tab. 1. Because 0.04 of them are not tagged in the Gd-LS,
these background events cannot be eliminated.  The
uncertainties of the neutron background from the PMTs in Tab.\ref{tab:bg} are from the binned neutron spectra in Ref.\cite{Mei2009}. But the
neutron background errors from the statistical fluctuation (their
relative errors are less than 1$\%$) are too small to be taken into
account.
\subsubsection{Neutrons from copper vessels}
In the copper vessels, neutrons are produced by the U and Th
contaminations and emitted with their average energy of 0.81
MeV\cite{Mei2009}. Their total volume is about $7.2\times10^4$
$cm^3$. The radioactive impurities Th can be reduced to
$2.5\times10^{-4}$ ppb in some copper samples\cite{Martin2009}.
 If we conservatively assume a 0.001 ppb U/Th
concentrations in the copper material\cite{Exo200_2007}, a rate of
one neutron emitted per $4\times10^4$ $cm^3$ per year is
estimated\cite{GdWater2008}. Consequently, there are 1.8 neutrons
produced by the all copper vessels per year. A simulated sample of 1.8$\times$$10^5$ events is used to study this neutron background.
These events generated isotropically are uniformly distributed in the copper material of the copper vessels.
\par
The simulation result is summarized in Tab.\ref{tab:bg}. 0.03
neutron events/(ton$\cdot$yr) reach the CsI(Na) target and their
energy deposition falls in the same range as that of the WIMP
interactions (see Tab.\ref{tab:bg}). As 0.001 of them are not tagged
in the Gd-LS, these background events cannot be eliminated. The
uncertainty of the neutron background from the copper vessels in Tab.\ref{tab:bg} are
from the binned neutron spectra in the Ref.\cite{Mei2009}. But the
neutron background errors from the statistical fluctuation (their
relative errors are less than 1$\%$) are too small to be taken into
account.
\subsubsection{Neutrons from other components}
The U and Th contaminations in other detector components also
contribute to the neutron background in our experiment setup.
Neutrons from the Aluminum reflectors are emitted with the average
energy of 1.96 MeV\cite{Mei2009}. The U and Th contaminations in the
carbon material are considered as the only neutron source in the
Gd-LS/LS. Neutrons from the Gd-LS/LS are emitted with the average
energy of 5.23 MeV\cite{Mei2009}. The U and Th contaminations in the
$\rm{SiO_2}$ material are considered as the only neutron source in the
PMTs in the oil. Neutrons from PMTs are emitted with the average
energy of 2.68 MeV\cite{Mei2009}. The U and Th contaminations in the
iron material are considered as the only neutron source in the
stainless steel tanks. Neutrons from the stainless steel tanks are
emitted with the average energy of 1.55 MeV\cite{Mei2009}. We
evaluated the neutron background from the above components using the
Geant4 simulation. All the nuclear recoils in the CsI(Na) target,
which fall in the same range as that of the WIMP interactions, are
tagged. The neutron background from these components can be ignored.
The simulated samples that amount to 1000 years of data taking are used to evaluate these neutron backgrounds.
\subsection{Neutron background from natural radioactivity in the surrounding rock}
In the surrounding rock, almost all the neutrons due to natural
radioactivity are below 10 MeV \cite{GdWater2008, Carson2004}. Water
can be used for shielding neutrons effectively, especially in the
low energy range of less than 10 MeV \cite{Carmona2004}. The WIMP
detectors are surrounded by about 2.5 meters of water and more than
1 meter of Gd-LS/LS, so these shields can reduce the neutron
contamination from the radioactivities to a negligible level.
\subsection{Neutron background due to cosmic muons}
Neutrons produced by cosmic muon interactions constitute an
important background component for dark matter searches. These
neutrons with a hard energy spectrum extending to several GeV
energies, are able to travel far from produced vertices.
\par
\begin{figure}
 \centering
 \includegraphics[width=0.8\textwidth]{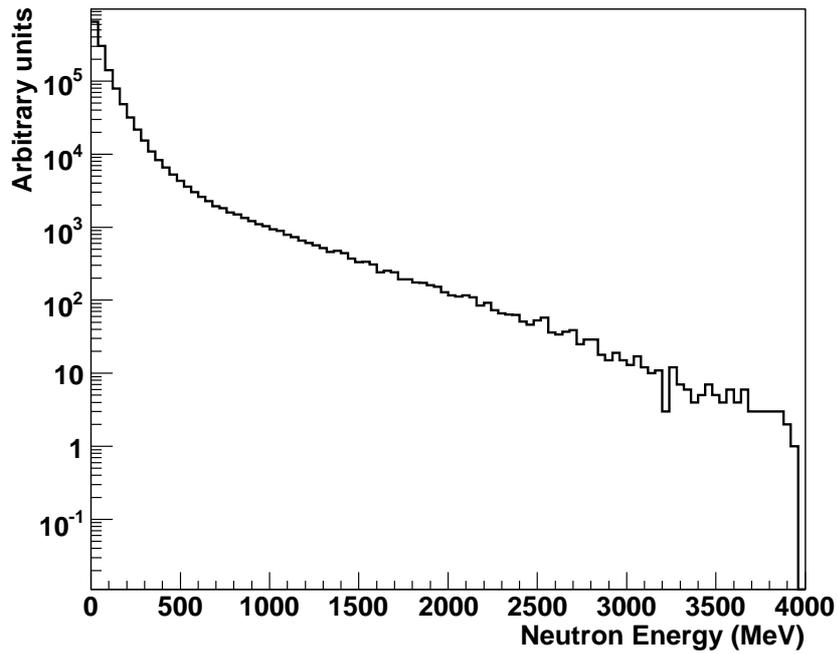}
 \caption{The energy spectrum of cosmogenic neutrons at depth of 910 m.w.e.}
 \label{fig:energy}
\end{figure}

The total cosmogenic neutron flux at a depth of 910 m.w.e. is
evaluated by a function of the depth for a site with a flat rock
overburden \cite{Mei2006}, and it is 1.31$\times$$10^{-7}$
$cm^{-2}s^{-1}$ . The energy spectrum (see Fig.\ref{fig:energy}) and
angular distribution of these neutrons are evaluated at the depth of
910 m.w.e. by the method in \cite{Mei2006, Wang2001}. The neutrons
with the specified energy and angular distributions are sampled on
the surface of the cavern, and the neutron interactions with the
detector are simulated with the Geant4 package. A simulated sample
of 1.25$\times$$10^9$ events is used to study in this neutron background.
\par
Tab.\ref{tab:bg}
shows that 2.5 neutron events/(ton$\cdot$yr) reach the CsI(Na) target
and their energy deposition falls in the same range as that of the WIMP
interactions. 0.28 of them are not tagged by the Gd-LS/LS. Muon veto
systems can tag muons very effectively, thereby most cosmogenic
neutrons can be rejected. In the Daya Bay experiment, the
contamination level can even be reduced by a factor of more than
30\cite{DayabayProp2007}. We assume the neutron
contamination level from cosmic muons decreases by a factor of 30
using a muon veto system. This could lead to the decrease of
cosmogenic neutron contamination to 0.01 events/(ton$\cdot$yr). The
uncertainties of the cosmogenic neutron background in
Tab.\ref{tab:bg} are from the statistical fluctuation.
\section{Rough estimation of other background}
Besides neutron background, other background events are mainly from reactor
neutrino events and electron recoils in the experimental design in
the present paper. The contamination caused by electron recoils
consists of bulk electron recoil events and surface events.
\subsection{Contamination due to reactor neutrino events}
Since neutrino detectors are fairly close to nuclear reactors (about
2 kilometers away) in reactor neutrino experiments, a large number
of reactor neutrinos will pass through the detectors, and nuclear
recoils will be produced by neutrino elastic scattering off target
nucleus in the WIMP detectors. Although neutrinos may be a source of
background for dark matter searches, they can be reduced to a
negligible level by setting the recoil energy threshold of 10
keV\cite{Jocely2007}. Besides, nuclear recoils may also be produced
by low energy neutrons produced by the inverse $\beta$-decay
reaction $\bar{\nu_{e}}+p\rightarrow e^{+}+n$. But their kinetic
energies are almost below 100 keV\cite{chooz2003}, and their maximum
energy deposition in the WIMP detectors is as large as a few keV.
Thus the neutron contamination can be reduced to a negligible level
by the energy threshold of 10 keV.
\subsection{Contamination due to bulk electron recoils}
The intenal sources of electron recoils are mainly caused by the radioisotopes of $\rm{^{137}Cs}$ and
$\rm{^{87}Rb}$ in the CsI crystals, and the external ones are mainly from the radioisotopes of $\rm{^{238}U}$, $\rm{^{232}Th}$ and $\rm{^{40}K}$
 in the 3" PMTs inside the copper vessels.
Here we assume that the concentrations of the $\rm{^{137}Cs}$ and $\rm{^{87}Rb}$ in the CsI crystals are about 2 mBq/kg and 1 ppb\cite{kims2003, kims2007}, respectively.
Then their background rates will be about 1.7 counts/keV/kg/day (cpd) in the region 10keV\cite{kims2007}.
While the concentrations of U/Th/K are about 78, 25 and 504 mBq/PMT, respectively\cite{kims2007}, and their background rates will be about 0.5 cpd in
the region 10keV.  Considering the fact that the rejection power
against electron recoils can reach O($10^7$)\cite{slz}, we
roughly estimate that the bulk electron recoil contamination due to
the internal and external sources of the CsI(Na) crystals is about 0.9 events/(ton$\cdot$yr).
\subsection{Contamination due to surface events}
The surface events for CsI(Na) crystals are caused by the deliquescence which reduces the $\rm{Na^+}$ concentration on
the crystal surface\cite{slz}. Signals produced by $\alpha$ particles from the progenies of $\rm{^{222}Rn}$ in the air
on the "old" surface of the crystal mimic nuclear recoils seriously. The surface events can be prevented from by
avoiding the deliquescence of the crystal surface. For example, the copper vessels are filled with dry nitrogen gas\cite{kims2007}, and thus
there is no air between the copper vessels and CsI(Na) crystals. Hence the surface event contamination can be reduced to a negligible level
by the above methods.

\section{Discussion and conclusion}
Neutron background can be effectively suppressed neutrino
detectors used as a neutron veto system in direct dark matter
searches. Tab.\ref{tab:bg} shows the total neutron contamination are
0.05 events/(ton$\cdot$yr). And compared to Ref.\cite{GdLS2010}, it
is reduced by a factor of about 6. If the electron recoils contaminations are considered, the total background
amounts to about 1 events/(ton$\cdot$yr).
\par
According to our work, the neutron background is
mainly from the PMTs in the copper vessels in this configuration with the
CsI(Na) targets. After
finishing a precision measurement of the neutrino mixing angle
$\theta_{13}$, we can utilize the existing experiment hall and
neutrino detectors. This will not only save substantial cost and
time for direct dark matter searches, but the neutron background
could also decrease to O(0.01) events per year per tonne of CsI(Na) in
the case of the Daya Bay experiment.  According to
Ref.\cite{Mei2006}, The neutron fluxes in the RENO (in an
underground laboratory with a depth of 450 m.w.e.), Double CHOOZ (in
an underground laboratory with a depth of 300 m.w.e.)
experiments\cite{reno,dc} are respectively about 5 and 3 times more than that of
the Daya Bay experiment. Their neutron backgrounds are roughly
estimated to be about 0.1 events/(ton$\cdot$yr), if their detector
configurations are the same as the one described above. The neutron contamination
is one order of magnitude smaller than the electron recoil contamination,
so neutron contamination can be ignored in this detector configuration.
\par
To evaluate the detector capability of directly detecting dark
matter, we assume a standard dark matter galactic
halo\cite{Lewin1996}, an energy resolution that amounts to 25$\%$
for the energy range of interest and 40$\%$ nuclear recoil
acceptance\cite{kims}.
\par
\begin{figure}
 \centering
 \includegraphics[width=0.8\textwidth]{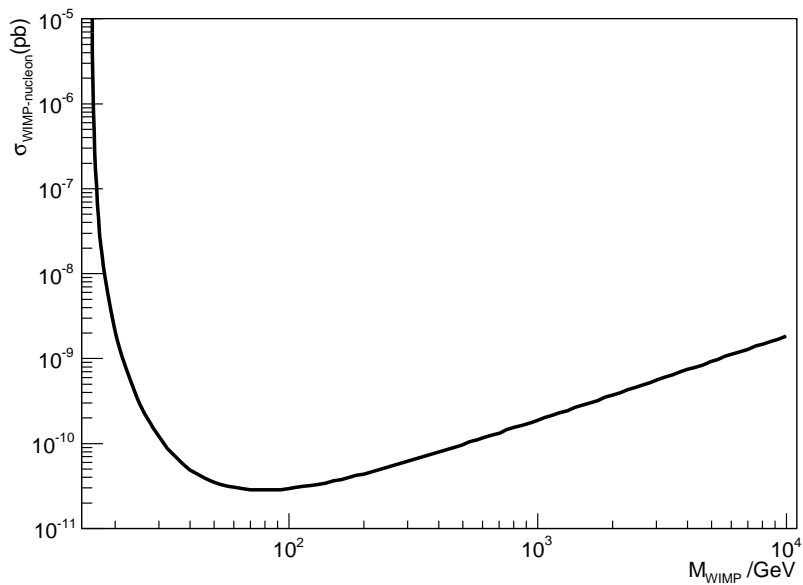}
 \caption{We calculate the sensitivity to spin-independent
WIMP-nucleon elastic scattering assuming an exposure of one tonne
$\times$ year. The calculation shows this exposure could reach a
cross-section of about 3$\times$$10^{-11}pb$ at the 90$\%$
confidence level. The tool from Ref.\cite{limits} has been used. }
 \label{fig:limit}
\end{figure}

If no signals are significantly observed, a sensitivity to WIMP-nucleon
spin-independent elastic scattering can be calculated via the same
method as Ref.\cite{GR1998}. Our calculation shows that an exposure
of one tonne $\times$ year could reach a cross-section of about
3$\times$$10^{-11}pb$ at the 90$\%$ confidence level (see
Fig.\ref{fig:limit}).
\section{Acknowledgements}
This work is supported in part by the National Natural Science Foundation
of China (NSFC) under contract No. 11235006 and the science fund of Fujian
University of Technology under contract No. GY-Z13114.
\par

\newpage
\begin{table}

  \centering
  \begin{tabular}{|c|c|c|}
  \hline
  & 20keV<$E_{recoil}$<100keV & Not Tagged\\
  \hline
  PMTs in copper vessels & 2.9$\pm$0.46 & 0.04$\pm$0.006 \\
  \hline
  copper vessels & 0.03$\pm$0.005 & 0.001$\pm$0.0002\\
  \hline
  cosmic muons & 2.5$\pm$0.45  & 0.28$\pm$0.15\\
  \hline
  muon veto & 0.08$\pm$0.08 & 0.01$\pm$0.028\\
  \hline
  total(muon veto) & 3.0$\pm$0.47 & 0.05$\pm$0.03\\
  \hline
  \end{tabular}
  \caption{Estimation of neutron background from different sources for
an underground laboratory at a depth of 910 m.w.e\@. The column
labeled "20keV<$E_{recoil}$<100keV" identifies the number of
neutrons whose energy deposition in the CsI(Na) target is in the same range as
WIMP interactions. The column labeled "Not Tagged" identifies the
number of neutrons which are misidentified as WIMP signatures (their
energy deposition in the CsI(Na) target is in the same range as WIMP
interactions while their recoil energies in the Gd-LS/LS are less
than the energy threshold of 1 MeV). The row labeled "copper vessel"
identifies the number of neutrons from the copper vessels.  The row labeled "PMTs in copper vessel"
identifies the number of neutrons from the PMTs in the copper vessels. The row
labeled "cosmic muons" identifies the number of cosmogenic neutrons
in the case of not using the muon veto system. The row labeled "muon
veto" identifies the number of cosmogenic neutrons in the case of
using the muon veto system. We assume that neutron contamination
level from cosmic muons decreases by a factor of 30 using a muon
veto system. Only the total background in the case of using the muon
veto system is listed in this table. The terms after $\pm$ are
errors.}
\label{tab:bg}
\end{table}

\end{document}